# The case for emulating insect brains using anatomical "wiring diagrams" equipped with biophysical models of neuronal activity

Logan T. Collins[1,*]

[1]Department of Psychology and Neuroscience, University of Colorado, Boulder

*Corresponding author:

Logan T. Collins

2860 Wilderness Place, Boulder, CO, 80301

email: loco1544@colorado.edu

ORCID: 0000-0003-1239-4310

**ABSTRACT:** Developing whole-brain emulation (WBE) technology would provide immense benefits across neuroscience, biomedicine, artificial intelligence, and robotics. At this time, constructing a simulated human brain lacks feasibility due to limited experimental data and limited computational resources. However, I suggest that progress towards this goal might be accelerated by working towards an intermediate objective, namely insect brain emulation (IBE). More specifically, this would entail creating biologically realistic simulations of entire insect nervous systems along with more approximate simulations of non-neuronal insect physiology to make "virtual insects." I argue that this could be realistically achievable within the next 20 years. I propose that developing emulations of insect brains will galvanize the global community of scientists, businesspeople, and policymakers towards pursuing the loftier goal of emulating the human brain. By demonstrating that WBE is possible via IBE, simulating mammalian brains and eventually the human brain may no longer be viewed as too radically ambitious to deserve substantial funding and resources. Furthermore, IBE will facilitate dramatic advances in cognitive neuroscience, artificial intelligence, and robotics through studies performed using virtual insects.

**Introduction**

*In silico* brain emulation represents a much sought-after dream within the field of computational neuroscience (Jordan et al., 2018; Koene, 2013; Markram, 2006; Markram et al., 2015). WBE would provide a platform for extremely rapid and precise investigations into cognition. Virtual activation and repression of neuronal subcompartments, individual neurons, and populations of neurons might be carried out to test the functional interdependence of anatomically distinct regions across multiple scales. Mathematical analyses performed on simulated neural activity could help uncover the mechanisms of neural circuits (Marder & Taylor, 2011; Park & Friston, 2013; Rowat, 2007). Although the resulting data would still be a model of the biological reality, incorporating appropriate levels of detail (i.e. synaptic connectivity and neuronal morphology) into the simulation may provide sufficient accuracy to replicate biological information processing (Koene, 2012) and so provide valuable insights regarding cognition and behavior.

Despite its status as an intermediary step towards the goal of emulating the human brain, IBE has immense promise for elucidating a more generalized understanding of cognitive processes and disorders since insects exhibit remarkably complex behaviors for their apparent simplicity. Even with its fairly small brain of 135,000 neurons (Alivisatos et al., 2012), *Drosophila melanogaster* integrates multiple streams of sensory information and exhibits decision making which goes beyond instinctually programmed responses (Gorostiza, 2018). In addition, *Drosophila* has demonstrated success as an animal model for intellectual disability and Alzheimer's disease, highlighting the utility of insects in biomedicine (Chakraborty et al., 2011; van der Voet, Nijhof, Oortveld, & Schenck, 2014). Honeybees demonstrate even more advanced cognitive abilities (Menzel, 2012). They show numerical cognition or "counting" (Pahl, Si, & Zhang, 2013), long-term memory on the scale of months (Menzel, 1999), and social communication regarding the spatial location of food through the "bee dance" (Menzel et al., 2011). As such, gaining a thorough understanding of insect cognitive machinery through IBE would represent an enormously valuable advance towards understanding neurological function and dysfunction.

IBE also has numerous applications in artificial intelligence and robotics since many insects exhibit high-level decision making and social communication. By contrast, most current artificial intelligence systems are "savants" that learn to perform certain tasks efficaciously but lack the agility of biological intelligence when dealing with the myriad challenges found in navigating a complicated world. Artificial intelligence can play games like chess and Go (Silver et al., 2017), accurately diagnose diseases based on symptomatic criteria (Yu, Beam, & Kohane, 2018), recognize and classify images containing particular objects (Akata, Perronnin, Harchaoui, & Schmid, 2014), and find elusive patterns within scientific data (Jimenez & Landgrebe, 1998). However, more humanlike artificial intelligence which can perform a myriad of distinct tasks as necessitated by the environment has proven challenging (Petrović, 2018).

IBE may accelerate the development of stronger artificial intelligence by enabling rapid and detailed studies of the neural computations related to versatile and complex insect behaviors. Furthermore, the immense diversity of macroscopically visible adaptations found among insects likely coincides with a similarly vast array of untapped cognitive mechanisms that may serve as the basis for biomimetic artificial intelligence and robotics. Understanding such mechanisms and their systems-level interactions could facilitate design of substantially more adaptable artificially intelligent agents. Even without complete mechanistic understanding, the circuits of insect intelligence could be borrowed and incorporated into synthetic cognitive agents. It should be noted that this possibility may partly depend on the modularity of insect brain structures. There is some evidence for modularity within insect brains, though the data still indicate that modules exhibit substantial crosstalk (Menzel & Giurfa, 2001). Nonetheless, IBE would still open the door to an enormous wealth of evolutionarily validated cognitive tools for the field of artificial intelligence.

Roboticists often attempt to design robots that mimic the motor abilities of biological organisms, so investigations on how virtual insect nervous systems control motor actions could benefit the design of autonomous mechanical agents. Many robots already use insect locomotion as an inspiration, including ground-based robots (Lambrecht, Horchler, & Quinn, 2005; Lim, McCarthy, Shaw, Cole, & Barnes, 2006; Nguyen et al., 2018) and aerial robots (Y. Chen et al., 2017; Zou, Zhang, & Zhang, 2016). In some cases, biomimetic robots have borrowed tools from insect cognition. Weiderman et al. investigated a neural circuit from dragonflies which facilitates tracking of visual targets and used this neural circuit to guide the design of a robot that follows moving objects (Wiederman, 2017). The successes of these efforts indicate that insect-inspired robotics could greatly benefit from the detailed computational understanding of insect sensorimotor circuits which may come from IBE.

**High-throughput structural mapping of insect connectomes**

IBE will necessitate powerful experimental tools for mapping insect brains at a level which resolves dendritic morphologies and synaptic contacts. Electron microscopy (EM), expansion microscopy (ExM), and x-ray microtomography (XRM) possess promise for attacking this challenge. It should be noted that functional methods may also complement structural imaging. For instance, Franconville et al. employed simultaneous optogenetic stimulation of presynaptic neurons and two-photon calcium imaging of possible postsynaptic neurons in the *Drosophila* central complex, establishing synaptic connectivity by examining downstream neuronal responses (Franconville, Beron, & Jayaraman, 2018). EM, ExM, and XRM, and other complementary techniques might allow mapping of insect connectomes in enough detail to enable IBE.

EM provides extremely high resolution but is a time-intensive technique even for small tissue volumes (Fig. 1A) (Denk, Briggman, & Helmstaedter, 2012; Helmstaedter et al., 2013; Marx, 2013). Nonetheless, EM has made major strides towards reconstructing insect connectomes. Using a customized high-throughput serial section transmission electron microscopy (ssTEM) platform, Zheng et al. acquired image data for the entire *Drosophila* brain (Zheng et al., 2018). Furthermore, a first-draft segmentation of the dataset has been carried out using a flood-filling network algorithm (Li et al., 2019). Since the algorithm was optimized to minimize fusion of distinct neurons (which is often especially problematic), the segmentation does contain numerous locations where single neurons were erroneously split into multiple segments. However, these split errors can undergo manual correction an order of magnitude more rapidly than manual skeletonization. Despite the customized EM platform used by Zheng et al., the method remains time-consuming. The authors reported that each 40 nm slice of *Drosophila* tissue took about seven minutes to image and that they successfully imaged 7,050 slices. Ignoring any possible interruptions, this means that the process took more than a month to complete. While this timescale is still impressive compared with other EM efforts, it might be challenging to scale the technique for the substantially larger brains of honeybees and other more complex insect species. EM is also unlikely to be the most efficient method for comparative connectomic studies between insect specimens in which many insect brains would undergo imaging. Even so, the dataset from this study represents an important step towards the construction of a virtual *Drosophila* and may pave the way for further connectomics efforts to facilitate the development of IBEs.

Several investigations have employed EM methods to reconstruct and more thoroughly annotate subsets of the *Drosophila* brain. Takemura et al. used focused ion-beam milling scanning electron microscopy (FIBSEM) to image all of the neurons and synapses within the fly's mushroom body α lobe (Takemura, Aso, et al., 2017). With these connectomic data, insights around dopaminergic modulation, memory formation, and parallel processing within the mushroom body were revealed. Another study employed FIBSEM to image a portion of the *Drosophila* optic lobe which included circuits related to motion detection (Takemura, Nern, et al., 2017). This anatomical reconstruction uncovered likely mechanisms for *Drosophila*'s motion detection which were previously unclear. Eichler et al. utilized ssTEM to reconstruct a connectome of the larval *Drosophila* mushroom body and comprehensively describe its circuit motifs (Eichler et al., 2017). The connectivity data were also used to build a computational model which illuminated the circuit's mechanisms of associative learning. These investigations illustrate that anatomical interrogation of brain tissue can expose mechanistic insights.

ExM involves infusing neural tissue with a swellable polymer matrix that is equipped with fluorescent labels for desired biomolecules (Fig. 1B) (F. Chen, Tillberg, & Boyden, 2015). This allows linear tissue expansion, enlarging the sample without introducing excessive distortions and facilitating higher "effective resolution" when imaging. The ExM process also makes treated samples partially translucent, which enables nondestructive optical imaging of deep tissue structures. New three-dimensional fluorescence microscopy techniques such as light-sheet microscopy show promise for working in concert with ExM (Bürgers et al., 2019; Liu et al., 2018). Furthermore, Gao et al. demonstrated imaging of whole *Drosophila* brains using by combining tissue expansion with lattice light-sheet microscopy (Gao et al., 2019). The tissue expansion and the specialized optics of the custom-built lattice light-sheet microscope enabled imaging at a resolution of 60×60×90 nm and with acquisition times of 2-3 days. The fly brain was immunostained to facilitate imaging of dopaminergic neurons as well as all presynaptic active zones. ExM's efficacy for imaging large regions of tissue with high resolution (Murakami et al., 2018) indicates that it may continue to provide valuable contributions towards mapping insect connectomes.

XRM represents a powerful and largely unexploited tool for structural connectomics (Fig. 1C). XRM involves staining tissue samples with high-z contrast agents, rotating the samples while scanning with x-rays, and then computationally reconstructing three-dimensional images.

Although it has a lower resolution than EM, dendritic spines have still been shown to be visible in XRM images of brain tissue (Fonseca et al., 2018; Mizutani et al., 2010; Pacureanu et al., 2019). XRM is nondestructive, works on timescales of hours rather than months, and needs less computational resources than EM for three-dimensional reconstruction (Mizutani et al., 2016). It has been applied in human tissue samples to help understand neural circuits (Mizutani et al., 2010). Furthermore, XRM has successfully reconstructed a skeletonized version of a *Drosophila* brain hemisphere with a resolution of about 600-800 nm, highlighting its potential for imaging insect brains (Mizutani, Saiga, Takeuchi, Uesugi, & Suzuki, 2013). Much like EM, the technique is still limited in terms of the person hours required for tracing neuronal processes, though improved neural tracing software which operates in a fully automated fashion may ameliorate this problem (Acciai, Soda, & Iannello, 2016; Donohue & Ascoli, 2011). If this computational challenge is overcome, XRM could provide a platform for rapid imaging and reconstruction of insect connectomes.

While purely structural data allows for detailed biophysical modeling of isolated neurons, the types of synaptic coupling and other molecular features will be essential for describing the insect brain at the network level. Fortunately, the outlined tools can be adapted for the purpose of synapse classification. In some cases, EM possesses sufficient resolution to allow identification of excitatory and inhibitory synapses by observing their morphological characteristics (Kleinfeld et al., 2011). Expansion microscopy is compatible with immunohistochemistry and genetically encoded fluorescent markers (F. Chen et al., 2015). XRM may allow absorption-based tagging of synaptic features using contrast agents that have distinct electron densities (Handschuh, Beisser, Ruthensteiner, & Metscher, 2017). Immunohistochemical techniques may facilitate tagging of other molecular features besides synaptic biomarkers with either fluorophores or x-ray contrast agents. For instance, antibodies which react with unique biomarkers expressed by non-spiking neurons might be used to identify which cells are non-spiking. These techniques may facilitate the construction of more realistic models for IBE.

**Translating structural data to biophysical models**

I propose that to construct an effective IBE, detailed neuroanatomical data from the desired insect will need to be combined with conductance-based biophysical models of neurons or other models which carry out biologically realistic simulation of dendritic processing. The Human Brain

Project (HBP) has made strides towards goals similar to IBE, but this effort has not emphasized biologically accurate neural connectivity (Markram, 2006; Markram et al., 2015; Reimann et al., 2017). The Human Brain Project has instead created virtual cortical columns using known densities of distinct morphological cell types within cortical layers and modeled synaptic coupling by implementing "typical" connectivity patterns for the layer under consideration. This technique develops rough approximations of biological neuroanatomy and is unlikely to be suitable for making IBEs that accurately reproduce behavior *in silico*. Nonetheless, the Human Brain Project does use multicompartmental Hodgkin-Huxley-type models at the network scale and so may provide valuable lessons on the practice of modeling detailed neurophysiology within large neuronal ensembles. In another investigation which may have relevance for biologically realistic neuronal simulation, Ujfalussy et al. developed a hierarchical linear-nonlinear model (hLN) to represent the nonlinear processing of dendritic arbors (Ujfalussy, Makara, Lengyel, & Branco, 2018). As such, the linear-nonlinear subunits corresponded to portions of the dendritic tree and were linked together accordingly. The parameters of the subunits were fitted to voltage and synaptic input data from a highly realistic multicompartmental model. After fitting, the hLN model demonstrated highly similar activity compared to the multicompartmental model. In future applications, hLN models could be more easily fit to *in vivo* data and may provide more easily interpretable functional descriptions of neuronal activity as compared to biophysical models. I suggest that IBE will require biologically realistic simulations with regards to both connectivity and dendritic processing.

Past investigations into insect computational neuroscience provide precedent for larger-scale efforts. Günay et al. used a multicompartmental Hodgkin-Huxley model to simulate a reconstructed motoneuron from *Drosophila* (Günay et al., 2015). In this way, the precise anatomical locations of distal ionic currents were predicted, demonstrating that the multicompartmental approach grants predictive accuracy. MaBouDi et al. constructed a spiking neural network emulation of an antennal lobe pathway associated with olfactory learning in honeybees (MaBouDi, Shimazaki, Giurfa, & Chittka, 2017). A spike-timing dependent plasticity model for the synapses between the antennal lobe neurons and outgoing projection neurons along with a model of octopaminergic modulation were used to simulate an olfactory discrimination process. Although this simulation used a leaky integrate-and-fire model rather than a multicompartmental Hodgkin-Huxley-type model, it still emphasized biological accuracy at the circuit level and demonstrated results that were consistent with the positive olfactory discrimination behaviors of bees, supporting the idea that biologically accurate modeling facilitates the emergence of biologically realistic outcomes. In addition, Ardin et al. built a rough model of the mushroom body of the desert ant *Cataglyphis velox* and used this simulation to control the navigation of agents in a virtual environment (Ardin, Peng, Mangan, Lagogiannis, & Webb, 2016). The model carried out learning using an input layer of visual projection neurons, a Kenyon cell layer for sparse encoding of visual inputs, and an output extrinsic neuron. The network used Izhikevich neurons equipped with a spike-timing dependent plasticity model and was trained using image data from a chosen navigational path through the virtual environment. When an image from this route was paired with a pattern of Kenyon cell activation, the synaptic weights between those Kenyon cells and the extrinsic neuron were greatly decreased. After training, the network was able to choose correct directions by following the minimum of extrinsic neuron activation. These studies show that even limited information on a neuronal circuit can facilitate creation of successful models, indicating that more complete information may allow for highly realistic recapitulation of insect behavior.

      Efforts towards developing larger-scale models of insect cognition have also started to emerge, providing a foundation for future IBE. In the Flysim project, Huang et al. used image data from the FlyCircuit database to develop a rough *Drosophila* whole-brain simulation (Huang et al., 2019). More than 20,000 reconstructed neurons from the FlyCircuit database were registered into a standard *Drosophila* brain space, allowing estimation of synapse locations using an algorithm which took both distance and number of contact points into account. Neuronal polarity (i.e. which end represents the dendritic arbor and which end represents the axonal projections) was estimated using a machine learning algorithm, neurotransmitter type was derived from the FlyCircuit database, and electrophysiological parameters were defined according to literature values. Leaky integrate-and-fire neurons were employed along with synapse models that included short-term plasticity variables. This draft whole-brain simulation exhibited both greater dynamical stability against hyperactivity and more diverse neuronal firing patterns relative to a control simulation with randomized neuronal connectivity. These results are more closely aligned with activity found in biological brains than the results from the randomized version, illustrating that biologically realistic connectivity is an important factor in constructing brain simulations.

The Fruit Fly Brain Observatory (FFBO) is a set of software infrastructure tools which help support *Drosophila* brain emulation (Ukani et al., 2019). More specifically, the FFBO is an open-source platform that acts as a central repository for storing and comparing many different types of *Drosophila*-related data and as a suite of tools for constructing and working with computational models of *Drosophila* brain circuits. The FFBO's NeuroArch hub contains a wealth of data on neuronal morphology and location, connectivity, and biophysics. The FFBO incorporates Neurokernel, a software platform intended to facilitate the integration of many independently developed *Drosophila* simulations into a unified WBE (Givon & Lazar, 2016). Since a degree of anatomical modularity is found among the fly's neuropils, Neurokernel streamlines interconnection of models representing different neuropil modules. In this way, it may aid collaboration by allowing multiple research groups to contribute modular simulations towards the goal of emulating the entire fly brain. The FFBO also includes a graphical user interface and a natural language query interface to help users navigate the system. By organizing these tools in a centralized fashion, the FFBO acts as a powerful starting point for translating insect brain data into emulations.

Existing models of the insect central complex may help instruct efforts towards IBE. Kakaria and de Bivort used light microscopy datasets to guide the connectivity of a spiking model of *Drosophila*'s protocerebral bridge and ellipsoid body (Kakaria & de Bivort, 2017). These structures have been demonstrated to encode the fly's direction of movement during navigation (Seelig & Jayaraman, 2015). The model behaved as an attractor network, successfully mimicking responses found in its biological counterpart (Kakaria & de Bivort, 2017). Le Moël et al. designed anatomically-based central complex neural circuit models connected to agents within a virtual environment. These models offered possible explanations for insect navigational behaviors including memory-directed movement, memory recalibration when food is moved from its previous spatial location, novel shortcutting between remembered food locations, and minimizing overall travel distance during complex foraging missions (Le Moël, Stone, Lihoreau, Wystrach, & Webb, 2019). In addition, Givon et al. used the FFBO as a platform to develop a model of the fly's central complex circuitry (Givon, Lazar, & Yeh, 2017). Known central complex neuron morphologies were loaded into the NeuroArch database and spatial proximity between presynaptic and postsynaptic neurites was employed to algorithmically infer the locations of synapses. Visual stimulus data were fed through a receptive field model to provide inputs to simulated neurons located in the bulb microglomeruli of the central complex. With this setup, simulations of wild type and mutant versions of the central complex's neural circuits were compared, generating insights on the region's mechanisms of operation. The models implemented in these studies could provide valuable strategies for broader IBE research.

Tschopp et al. demonstrated another potentially useful tool for supporting IBE in the form of a connectome-based simulation of the *Drosophila* medulla and lamina which resulted in automatic emergence of orientation and direction selectivity properties (Tschopp, Reiser, & Turaga, 2018). Data from EM optic lobe reconstructions (Takemura et al., 2013) were used to build a simplified network of linear-nonlinear point neurons organized into a repeating hexagonal lattice. Synaptic weights were initialized as proportional to the number of biological synapses each neuron received in the EM reconstructions. The network was trained using a video-based input dataset with the goal of object tracking. Despite the dramatic simplifications of this model relative to its biological counterpart, it exhibited spontaneous orientation selectivity and direction selectivity after training facilitated fine-tuning of the weights. Because this strategy allowed prediction of functional properties from structural data, similar methods might be applicable to future work in converting imaging data to functional simulations of insect brains.

Although IBE will require detailed whole-brain data from insects, some simplifications might be possible while still maintaining strong biological realism. Despite neglecting molecular details, multicompartmental conductance-based models are highly predictive of biological neural activity (Fig. 2A-B) (Herz, Gollisch, Machens, & Jaeger, 2006). I contend that multicompartmental models with modifications for emulating synaptic potentiation, non-canonical electrophysiological influences (i.e. dendritic calcium spikes and NMDA spikes), chemical signaling, and glial modulation may produce sufficient biological accuracy. These simulations may not need to follow the dynamics of individual biomolecules. For synaptic potentiation, models may take the form of spike-timing dependent plasticity functions equipped with terms that account for cooperativity among spike inputs (Rabinovich, Varona, Selverston, & Abarbanel, 2006; Sjöström, Rancz, Roth, & Häusser, 2008). The strong correlation between dendritic spine volume and synaptic efficacy could be leveraged to infer the strengths of synapses from structural data (Bartol Jr et al., 2015). Non-canonical electrophysiology might be incorporated into existing multicompartmental models (Destexhe, Contreras, Steriade, Sejnowski, & Huguenard, 1996; Holcman & Yuste, 2015). Glial modulation and other chemical signaling processes could be simulated using reaction-diffusion models that describe spatiotemporally dependent concentrations of various signaling molecules (McDougal, Hines, & Lytton, 2013). Neuronal connectivity can be computationally described using directed adjacency matrices, a graph theoretic technique (Bullmore & Sporns, 2009). Iterative comparison of IBEs to biological insects and refinement of models will be essential for achieving biological accuracy. Using these kinds of methods, virtual insect behavior may demonstrate close resemblance to the behavior of biological insects.

**Emulating non-neuronal physiology**

To create IBEs that provide meaningful insights regarding the connection between brain function and behavior, models of non-neuronal insect physiology will also require implementation. Fortunately, these processes may necessitate less detailed modeling to achieve biological realism. The simulated *C. elegans* created by Palyanov et al. provides evidence that such a simplification could be reasonable (Palyanov, Khayrulin, Larson, & Dibert, 2012). Their emulation of the worm included a neuromuscular system in which the musculature was approximated using spring constructs linked to appropriate points on a wireframe body. Even with this very rough model, wormlike movements were observed in the virtual *C. elegans*. As such, modeling non-neuronal physiology like that of the musculature might be feasible without the single-cell resolution mapping which is more important for emulating nervous systems.

The sensory organs are another important type of non-neuronal physiology to consider for IBE. Sensory organs use sophisticated mechanisms to transduce sensory information and will necessitate similarly sophisticated models. Nonetheless, progress has been made towards developing models that reproduce the dynamics of sensory operations. Clemens et al. developed models for audition in crickets using signal processing techniques to fit auditory data to electrophysiological responses (Clemens & Hennig, 2013; Clemens, Wohlgemuth, & Ronacher, 2012). Although these models are phenomenological rather than biophysical, they have been shown to accurately translate auditory information into neuronal activity. Likewise, there are well-established models for insect vision. The Reichardt detector describes motion detection in the context of the ommatidia, the optical units found on insect eyes (Borst, 2007; Reichardt, 1987). It compares luminance values at two locations and uses a temporal delay at one of the locations to facilitate motion detection. While the Reichardt detector does not comprehensively describe insect photoreception, it demonstrates that efficacious models can be developed for nontrivial aspects of visual processing in the insect eye. These examples demonstrate that insect sensory modalities are amenable to reasonably simple computational representations.

The insect endocrine system may represent one of the more difficult non-neuronal systems to describe for IBE. Insect endocrine systems, including those of the *Drosophila* and the honeybee, are fairly well-characterized (Bloch, Hazan, & Rafaeli, 2013; Even, Devaud, & Barron, 2012; Farooqui, 2012; Hauser, Cazzamali, Williamson, Blenau, & Grimmelikhuijzen, 2006; Orchard & Lange, 2012). However, endocrine physiology exhibits multiscale dynamics which range from molecular to whole-organism levels. Because of this, it will be essential to develop models that incorporate necessary mechanistic features of insect endocrine physiology via approximations which circumvent excess computational demands while still mediating reasonably accurate behavioral outcomes in silico. As hemolymph undergoes continuous flow throughout the insect haemocoel, the well-mixed system assumption is likely applicable to insect hormone transport. More difficulties may arise in modeling the complex modulatory effects of hormones on insect physiology, particularly when considering that the interplay of hormonal influences exhibits highly context-dependent properties (McKenna & O'Malley, 2002). High-throughput assays in which the context-dependent effects of many insect hormones are tested in a combinatorial fashion may aid in the development of entomological endocrine models. Endocrine systems present a challenge to IBE but not an insurmountable obstacle.

**Computational resources**

Though IBE models may allow simplifications relative to the biological systems they mimic, even a single IBE could demand a large amount of computational resources. While precisely estimating these requirements is challenging without more detailed specifications on model construction, I will speculate on some possibilities using floating-point operations per second (FLOPS) to describe the necessary levels of computational demands. Though FLOPS represent a rough metric, they still provide a reasonable "first guess." One method for generating such estimates involves multiplying the neuronal population size times the average input synapses per neuron times the mean spike frequency (Furber, Temple, & Brown, 2006). Assuming 1,000 input synapses per neuron on average with a mean spike frequency of 10 Hz, emulating the

*Drosophila* brain would require $10^9$ FLOPS and emulating the honeybee brain would require $10^{10}$ FLOPS.

However, this approach for estimating computational demands focuses on network-level processing and ignores the requirements of computation within individual neurons. Although such an approximation would be viable for an emulation that utilizes McCulloch-Pitts or integrate-and-fire neurons, these models are almost certainly too far simplified to possess sufficient biological realism. I will further multiply by a factor that describes the necessary resources to emulate each neuron using multicompartmental biophysical models. For a single neuron, such Hodgkin-Huxley-type models require around $1.2 \times 10^6$ FLOPS (Izhikevich, 2004). Taking the product of this factor with the network-based estimate, a *Drosophila* IBE would need $1.2 \times 10^{15}$ FLOPS and a honeybee brain would need $1.2 \times 10^{16}$ FLOPS. As non-neuronal physiology will likely require less resources than nervous systems, a virtual *Drosophila* probably would need less than $2 \times 10^{15}$ FLOPS and a virtual honeybee probably would require less than $2 \times 10^{16}$ FLOPS. These demands fall within the capabilities of the fastest existing supercomputers which operate at up to $2 \times 10^{17}$ FLOPS (Hines, 2018). As exascale supercomputers (which operate at speeds of $10^{18}$ FLOPS or higher) are planned for completion in the early 2020s (Lee & Amaro, 2018; Service, 2018), IBE represents a quite reasonable goal from a computational standpoint.

Application-specific hardware tools for computational neuroscience may further increase the accessibility of virtual insects. The organization of neuromorphic hardware more closely resembles neurobiology than the organization found in traditional circuitry (Indiveri et al., 2011). For this reason, neuromorphic hardware more efficiently runs emulations of neurobiological systems. The neuromorphic supercomputer SpiNNaker provides a large-scale example of neuromorphic hardware. The SpiNNaker hardware has emulated a cortical microcircuit of 80,000 leaky integrate-and-fire neurons and 300 million synapses (van Albada et al., 2018). With enough cores, the hardware could simulate up to a billion neurons in real time (Brown, Chad, Kamarudin, Dugan, & Furber, 2018). Field-programmable gate arrays (FPGAs) represent another type of promising neuromorphic architecture (Zjajo et al., 2018). FPGAs have demonstrated great potential for emulating biologically realistic neurons via multicompartmental Hodgkin-Huxley-type models than many other types of hardware. Neuromorphic computing may enhance the efficiency of IBE and so allow for simulations to be carried out at lower cost.

**Conclusions**

Beyond its immediate applications, IBE raises some important philosophical considerations. If an insect's behavior is successfully reproduced in a virtual setting using a biologically accurate brain emulation, the IBE may very well exhibit some form of consciousness. This possibility is supported by integrated information theory (IIT), an attempt to outline fundamental mathematical constraints that underlie the physical phenomena necessary for particular qualia to occur (Oizumi, Albantakis, & Tononi, 2014). IIT lends credence to substrate independence, the idea that any system with equivalent information processing will exhibit the same conscious experiences regardless of its substrate (i.e. neuromorphic silicon or biological neurons). As such, IBE provides an early opportunity to develop ethical guidelines for handling emulated minds. This will be vital if the human brain is eventually emulated in a nonbiological substrate. It would be far too easy to dismiss a human emulation as a nonhuman entity and then subject the emulation to experiments that cause suffering. Although substrate independence may or may not hold true, the possibility should be thoroughly investigated since a philosophical error could result in disturbing consequences. Nonetheless, I suggest that IBE and later emulations of human minds are worthwhile endeavors. If a proper ethical framework for handling WBE is developed, even human brain emulation could be carried out in a fashion that provides great benefits to the human species without harming the emulations themselves.

I argue that IBE is a feasible near-term goal (within 20 years) along the path to human WBE. Furthermore, IBE possesses numerous applications in biomedicine, artificial intelligence, and robotics. Much of the structural data required to construct virtual insects may come from technologies like EM, ExM, and XRM. Extended versions of these tools may also facilitate the acquisition of data regarding the types of neurotransmitters secreted from synapses. Anatomical "wiring diagrams" derived from such experimental data may enable the construction of detailed multicompartmental Hodgkin-Huxley-type models that exhibit biologically realistic dynamics. Emulation of non-neuronal physiologies may necessitate less detailed experimental data to build and less computational resources to run, but it will still require substantial research to develop. Based on the outlined models and the neuronal population sizes of insect brains, I suggest that the computational demands of IBEs may fall within the capacities of existing supercomputers and well within the capacities of upcoming supercomputers currently under construction. Neuromorphic hardware architectures may further increase the computational efficiency of IBE. Building an IBE represents a difficult endeavor, but I propose that it can be achieved given organized effort and multidisciplinary collaboration.

**ACKNOWLEDGEMENTS:**

The work was supported by the Arnold and Mabel Beckman Foundation under the Beckman Scholars Program. I thank Michael P. Saddoris for his constructive feedback on the manuscript.

Conflict of interest: The author declares no competing financial interests.

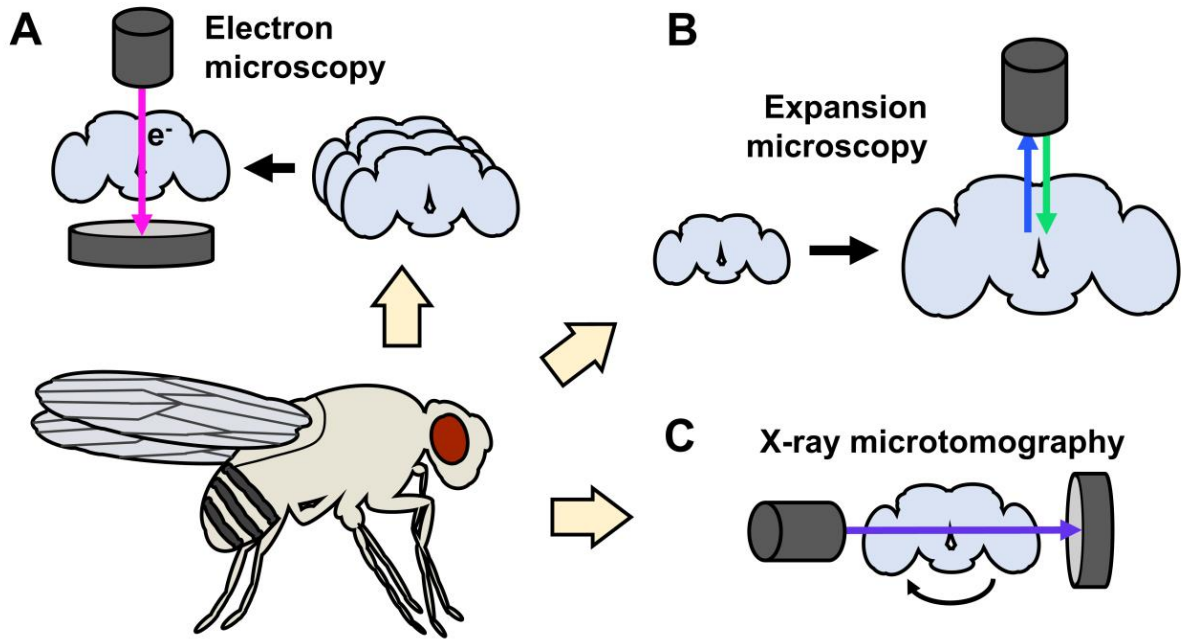

**Fig. 1** Techniques for structural mapping of insect brain tissue. **(a)** Electron microscopy (EM) requires preparation of ultrathin tissue slices. An electron beam is then passed through each slice and the resulting images are computationally stacked to obtain a 3D reconstruction with nanometer-scale resolution. Even with high-throughput automation, this technique is highly time-intensive. Nonetheless, EM remains the gold standard for connectomics due to its very high resolution. **(b)** Expansion microscopy (ExM) physically enlarges tissue via an infused polymer matrix (F. Chen et al., 2015). This facilitates "effective super-resolution imaging" using fluorescence microscopy. Emerging technologies like light-sheet microscopy may enable rapid 3D imaging of tissue volumes using ExM (Liu et al., 2018). **(c)** X-ray microtomography (XRM) passes x-rays through a sample positioned on a rotating stage and allows 3D reconstruction. For imaging soft tissue, XRM requires a high-Z contrast agent. Although XRM has lower resolution than EM, it can be performed much more efficiently for larger tissue volumes while still retaining a resolution sufficient to observe dendritic and axonal processes (Mizutani et al., 2011). In addition, XRM is a nondestructive method.

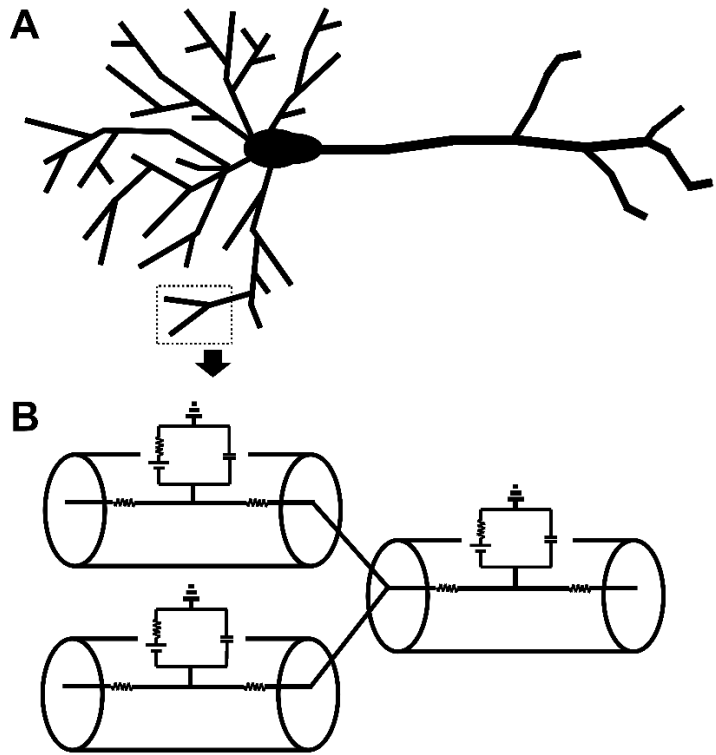

**Fig. 2** Multicompartmental Hodgkin-Huxley-type models. **(a)** Most biological neurons possess complex dendritic trees. These morphologies combine numerous excitatory and inhibitory postsynaptic potentials via a nonlinear process. As a result, a neuron's geometric constraints exert spatiotemporal control over membrane voltage propagation and dendritic computation. **(b)** Biologically realistic neurons can be modeled *in silico* using multicompartmental models that decompose the dendritic arbor into a set of interlinked segments and describe membrane voltage dynamics using Hodgkin-Huxley equations coupled to a partial differential equation known as the cable equation (Gerstner, Kistler, Naud, & Paninski, 2014).